\documentclass[12 pt]{article}
\usepackage[utf8x]{inputenc}
\usepackage{amsmath}
\usepackage{amsfonts}
\usepackage{amssymb}
\usepackage{empheq}
\usepackage[sort&compress,numbers]{natbib}
\usepackage{doi}
\hoffset= - 0.9in         

\newcommand{\p}{\partial}
\newcommand{\f}{\frac}
\newcommand{\s}{\sqrt}

\newcommand{\al}{\alpha}

\newcommand{\nn}{\nonumber}

\newcommand{\D}{\Delta}

\usepackage{hyperref}
\newtheorem{obs}{Observation}
\hypersetup{
  colorlinks   = true, 
  urlcolor     = blue, 
  linkcolor    = blue, 
  citecolor   = red 
}
\usepackage{hyperref}
\usepackage{tikz}
\usepackage{epigraph}
\usepackage{float}

\begin{document}
\title{On new exact conformal blocks and Nekrasov functions}
 \author{{Nikita Nemkov}\thanks{{\small {\it Moscow Institute of Physics and Technology (MIPT), Dolgoprudny, Russia}
 and {\it Institute for Theoretical and Experimental Physics (ITEP), Moscow, Russia}; nnemkov@gmail.com} }
\date{ }}
\date{\today}
\maketitle
\vspace{-5.0cm}

\begin{center}
 \hfill ITEP/TH-15/16\\
\end{center}

\vspace{3.5cm}
\begin{abstract}
Recently, an intriguing family of the one-point toric conformal blocks AGT related to the $\mathcal{N}=2^*\,\, SU(2)$ Nekrasov functions was discovered by M. Beccaria and G. Macorini. Members of the family are distinguished by having only finite amount of poles as functions of the intermediate dimension/v.e.v. in gauge theory. Another remarkable property is that these conformal blocks/Nekrasov functions can be found in closed form to all orders in the coupling expansion. In the present paper we use Zamolodchikov's recurrence equation to systematically account for these exceptional conformal blocks. We conjecture that the family is infinite-dimensional and describe the corresponding parameter set. We further apply the developed technique to demonstrate that the four-point spheric conformal blocks feature analogous exact expressions. We also study the modular transformations of the finite-pole blocks. 
\end{abstract}
\tableofcontents
\section{Introduction}
Conformal blocks \cite{BPZ} present a new class of special functions crucial to many questions in modern physics. In particular, due to the AGT relation \cite{AGT} they are equal to the partition functions of the $\Omega$-deformed SUSY gauge theories known as the Nekrasov functions \cite{Nekrasov:2003rj, Nekrasov:2002qd}. The set of conformal blocks known in closed explicit form is quite poor. Recently, in paper \cite{Beccaria:2016nnb} the authors enriched this class significantly by providing a number of examples when the toric conformal blocks are expressed via finite combinations of the Eisenstein series. They have found these cases by direct analysis of the Nekrasov functions at small instanton orders and then conjectured that certain cancellations hold in all orders. All these examples share a distinctive property: they correspond to the conformal blocks having only a finite amount of poles as functions of the intermediate dimension. In conformal field theory there is a wonderful formula describing analytic structure of conformal blocks as functions of the intermediate dimension, Zamolodchikov's formula. In this note we use it to give a new perspective on the family of exactly solvable cases found in \cite{Beccaria:2016nnb}. We make a plausible conjecture that the family of the finite-pole conformal blocks is infinite and describe the subspace of parameters where they appear. We further apply the developed technique to the four-point spheric conformal blocks and find exactly solvable cases there, too. Finally, we describe the modular transformations of the finite-pole blocks explicitly constructing the corresponding modular kernels.
\section{Zamolodchikov's formula}
Toric conformal block can be defined as the following trace
\begin{eqnarray}
B_\D(\D_e,c|q)=\operatorname{Tr}_\D\left(q^{L_0-\f{c}{24}}V_{\D_e}\right) \label{CB trace def}
\end{eqnarray}
with $c$ being the central charge, $q$ the torus complex structure parameter, $V_{\D_e}$ the primary field of the conformal dimension $\D_e$ (the external dimension), and $\D$ the dimension of the conformal family over which the trace is taken (the internal dimension). Conformal block is usually represented as a series in powers of $q$
\begin{eqnarray}
B_\D(\D_e,c|q)=q^{\D-\f{c}{24}}\sum_{n=0}^{\infty}q^nB_\D^n(\D_e) \label{CB q exp}
\end{eqnarray}
Coefficients $B^n_\D(\D_e)$ are known to be polynomials in $\Delta_e$ and rational functions in $\Delta$ and $c$. It is well known that for generic $c,\D_e$ conformal block has simple poles at the Kac zeros $\D_{r,s}$
\begin{eqnarray}
\D_{r,s}=\f{Q^2}{4}-\al_{r,s}^2,\quad \al_{r,s}=r\f{b}2+s\f{b^{-1}}2 \label{Kac zeroes}
\end{eqnarray}
where $r,s\ge1$ are natural numbers labeling the Kac zeros, while $Q$ and $b$ parametrize the central charge as 
\begin{eqnarray}
c=1+6Q^2,\qquad Q=b+b^{-1} \label{b def}
\end{eqnarray}
What is perhaps less known, the residues at the Kac zeros can be expressed via conformal blocks with specific internal dimensions. The exact relation reads
\begin{eqnarray}
B_\D(\D_e,c|q)=\chi_\D(c|q)+\sum_{r,s\ge1}\f{R_{r,s}(\D_e,c)}{\D-\D_{r,s}}q^{\D-\D_{r,s}}B_{\D_{r,-s}}(\D_e,c|q) \label{Zamolodchikov's formula}
\end{eqnarray}
where $\chi_\D(c|q)$ is the Virasoro character \footnote{$\eta(q)$ is the Dedekind eta function defined in appendix \ref{special functions}.}
\begin{eqnarray}
\chi_\D(c|q)=\f{q^{\D-\f{c-1}{24}}}{\eta(q)} \label{character}
\end{eqnarray}
and $q$-independent coefficients $R_{r,s}(\D_e,c)$ are given by
\begin{eqnarray}
R_{r,s}(\D_e,c)=\f{\al_{r,s}}{Q}\prod_{n=0}^{r-1}\prod_{m=0}^{s-1}\f{(\D_e-\D_{2n+1,2m+1})(\D_e-\D_{2n+1,-2m-1})}{\D'_{2n+1,2m+1}\D_{2n+1,-2m-1}} \label{residue}
\end{eqnarray}
Here $\Delta'_{2n+1,2m+1}=1$ if $n=m=0$ (note that $\D_{1,1}=0$) and $\Delta'_{2n+1,2m+1}=\Delta_{2n+1,2m+1}$ otherwise. 

We stress that the conformal block which in \eqref{Zamolodchikov's formula} enters the residue at $\D=\D_{r,s}$ has internal dimension $\D_{r,-s}\equiv \D_{r,s}+rs$ which is not degenerate for generic $c$.

Rewritten in terms of the coefficients $B^n(\D_e,c)$ from \eqref{CB q exp} formula \eqref{Zamolodchikov's formula} gives a recurrent relation allowing to compute conformal blocks order by order in $q$. Introducing the so-called elliptic block related to the one defined above by a character renormalization
\begin{equation}
H_{\D}(\D_e,c|q)=\f{B_\D(\D_e,c|q)}{\chi_\D(c|q)},\qquad H_\D(\D_e,c|q)=\sum_{n=0}^\infty q^nH^n_\D(\D_e,c) \label{elliptic CB def}
\end{equation}
formula \eqref{Zamolodchikov's formula} becomes
\begin{eqnarray}
H_\D(\D_e,c|q)=1+\sum_{rs\ge1}\f{R_{r,s}(\D_e,c)}{\D-\D_{r,s}}q^{rs}H_{\D_{r,-s}}(\D_e,c|q) \label{Zamolodchikov's elliptic}
\end{eqnarray}
or, in terms of the coefficients
\begin{eqnarray}
H^n_\D(\D_e,c)=\delta^{n}_{0}+\sum_{1\le rs\le n}\f{R_{r,s}(\D_e,c)}{\D-\D_{r,s}}H^{n-rs}_{\D_{r,-s}} \label{coeff recursion}
\end{eqnarray}
with $\delta^{n}_0$ being the Kronecker delta, the seed of the recursion. Further we mostly use the elliptic blocks \eqref{elliptic CB def} keeping in mind that they are simply related to the canonical ones \eqref{CB trace def}.

Equation \eqref{Zamolodchikov's elliptic} was proposed by Poghossian in \cite{Poghossian:2009mk} and proved in \cite{Hadasz:2009db}. The formula is a counterpart of the famous recursion relation found by Al. Zamolodchikov \cite{Zamolodchikov:Rec_c,Zamolodchikov:Rec_a} for spheric blocks. In the following we refer both to this formula and to the original formula for spheric blocks as to Zamolodchikov's formula. 
\section{Exact toric conformal blocks}
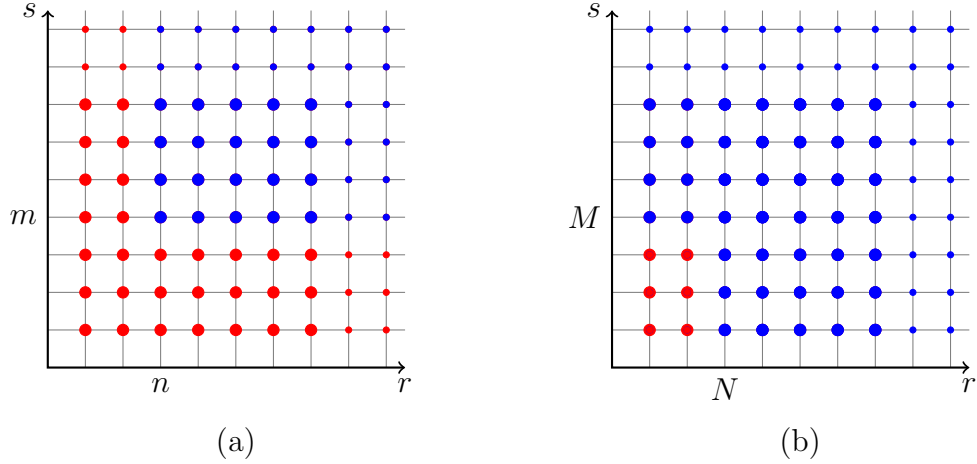
\begin{figure}
\begin{center}
\begin{tikzpicture}[scale=0.5]
\def\N{7}
\def\ND{9}
\def\NA{9.5}
\draw[gray, thin] (0,0) grid (\NA,\NA);  

\foreach \r in {1,...,\N}{
	\foreach \s in {1,...,\N}{
		\filldraw[color=red] (\r,\s) circle (0.15);
		}
	}
	
\foreach \r in {1,...,\ND}{
	\foreach \s in {1,...,\ND}{
		\filldraw[color=red] (\r,\s) circle (0.08);
		}
	}	
\foreach \r in {3,...,\N}{
	\foreach \s in {4,...,\N}{
		\filldraw[color=blue] (\r,\s) circle (0.15);
		}
	}
	
\foreach \r in {3,...,\ND}{
	\foreach \s in {4,...,\ND}{
		\filldraw[color=blue] (\r,\s) circle (0.08);
		}
	}

\draw[<->, thick] (\NA,0) -- (0,0) -- (0,\NA);

\node[below] at (3,0) {$n$};
\node[left] at (0,4) {$m$};
\node[left] at (0,\NA) {$s$};
\node[below] at (\NA,0) {$r$};
\node at (5,-2) {(a)};
\end{tikzpicture}
\qquad\qquad
\begin{tikzpicture}[scale=0.5]
\def\N{7}
\def\ND{9}
\def\NA{9.5}
\draw[gray, thin] (0,0) grid (\NA,\NA);  

\foreach \r in {1,...,\N}{
	\foreach \s in {1,...,\N}{
		\filldraw[color=red] (\r,\s) circle (0.15);
		}
	}
	
\foreach \r in {1,...,\ND}{
	\foreach \s in {1,...,\ND}{
		\filldraw[color=blue] (\r,\s) circle (0.08);
		}
	}	
\foreach \r in {1,...,2}{
	\foreach \s in {1,...,3}{
		\filldraw[color=red] (\r,\s) circle (0.08);
		}
	}	
\foreach \r in {3,...,\N}{
	\foreach \s in {1,...,\N}{
		\filldraw[color=blue] (\r,\s) circle (0.15);
		}
	}
\foreach \r in {3,...,\N}{
	\foreach \s in {1,...,\N}{
		\filldraw[color=blue] (\r,\s) circle (0.15);
		}
	}
	
\foreach \r in {1,...,\N}{
	\foreach \s in {4,...,\N}{
		\filldraw[color=blue] (\r,\s) circle (0.15);
		}
	}

\draw[<->, thick] (\NA,0) -- (0,0) -- (0,\NA);

\node[below] at (3,0) {$N$};
\node[left] at (0,4) {$M$};
\node[left] at (0,\NA) {$s$};
\node[below] at (\NA,0) {$r$};
\node at (5,-2) {(b)};
\end{tikzpicture}
\end{center}
\caption{Pole distributions in the $(r,s)$-plane: (a) in case when $R_{n,m}=0$ (b) in case when $R_{1,N}=R_{M,1}=0$. Poles with non-vanishing $R_{r,s}$ are depicted in red, the others in blue.}
\label{nm poles}
\end{figure}

Examples presented in \cite{Beccaria:2016nnb} suggest that there exist cases when conformal block contains finitely many poles in $\D$. It seems very natural to look at these examples from the standpoint provided by Zamolodchikov's formula \eqref{Zamolodchikov's elliptic}, the sum in which under these circumstances truncates at a finite term. The basic idea is plain. Coefficients $R_{r,s}(\D_e,c)$ \eqref{residue} have nested structure of zeros: once we choose $\D_e$ in such a way that $R_{n,m}(\D_e,c)$ for some $n$ and $m$ vanishes, so will $R_{r,s}(\D_e,c)$ for all $r\ge n,s\ge m$, see fig. \ref{nm poles} (a). However, vanishing of a single coefficient $R_{r,s}$ still leaves two infinite strips of poles located at $r\ge 1,1\le s < m$ and $1\le r<n,s\ge 1$. In order to obtain a finite amount of poles, it is necessary for a pair of coefficients $R_{N, 1}(\D_e,c), R_{1,M}(\D_e,c)$ to be zero for some $N,M$, see fig. \ref{nm poles} (b). As we shortly demonstrate, this requires tuning both, the external dimension and the central charge. Importantly, at the values of the central charge allowing for such truncation, the conformal blocks in the r.h.s. of \eqref{Zamolodchikov's elliptic} can be singular, and thus one has to carefully approach these special points. We first provide illustrations for a no-pole and a single-pole examples, and then discuss the general situation. 
\subsection{No poles}
The simplest case occurs when $N=M=1$. Then, adjusting parameters so that $R_{1,1}=0$ will also force all the other pole contributions to be absent in \eqref{Zamolodchikov's formula}. Explicitly
\begin{eqnarray}
R_{1,1}(\D_e,c)=\f{\al_{1,1}}{Q}\f{(\D_e-\D_{1,1})(\D_e-\D_{1,-1})}{\D'_{1,1}\D_{1,-1}}=\f12\D_e(\D_e-1)=0
\end{eqnarray} 
where we used $\al_{1,1}=Q/2,\, \D_{1,1}=0,\,\D_{1,-1}=1,\,\D'_{1,1}=1$. This allows us to choose $\D_e=0$ or $\D_e=1$. Since any $R_{r,s}(\D_e,c)$ with $r,s\ge1$ vanish upon this substitution, we obtain
\begin{eqnarray}
B_{\D}(0,c|q)=B_{\D}(1,c|q)=\chi_\D(c|q) \label{no poles cb}
\end{eqnarray}
with $\chi_\D(c|q)$ defined in \eqref{character}. For $\D_e=0$ this statement is obvious from the definition \eqref{CB trace def}. However, for $\D_e=1$ this is already non-trivial. Validity of these solutions regardless of the central charge value is a special feature of the no-pole case.
\subsection{Single pole} \label{single pole sec toric}
This subsection builds the toolbox and develops the intuition important throughout the rest of the paper. The single-pole case reveals all the main features appearing in the higher-pole solutions but with less computational effort.

To obtain a single-pole conformal block one has to require that $R_{1,2}$ and $R_{2,1}$ vanish \footnote{There is in fact another possibility to obtain a single-pole solution. It might be that several contributions are present in \eqref{Zamolodchikov's elliptic} but the poles of these contributions are the same for a given value of the central charge. We will return to this issue in subsection \ref{general situation}.}.  Explicitly this reads
\begin{align}
R_{1,2}=0:\quad \f{\al_{1,2}}{Q}\f{(\D_e-\D_{1,3})(\D_e-\D_{1,-3})(\D_e-\D_{1,1})(\D_e-\D_{1,-1})}{\D_{1,3}\D_{1,-3}\D'_{1,1}\D_{1,-1}}=0\\
R_{2,1}=0:\quad \f{\al_{2,1}}{Q}\f{(\D_e-\D_{3,1})(\D_e-\D_{3,-1})(\D_e-\D_{1,1})(\D_e-\D_{1,-1})}{\D_{3,1}\D_{3,-1}\D'_{1,1}\D_{1,-1}}=0
\end{align}
Choosing $\D_e=\D_{1,1}=0$ or $\D_e=\D_{1,-1}=1$ reduces the situation to the previous no-pole case. New possibilities are described by
\begin{align}
R_{1,2}=0:\quad \f{\al_{1,2}}{Q}\f{(\D_e-\D_{1,3})(\D_e-\D_{1,-3})}{\D_{1,3}\D_{1,-3}}=0 \label{R12=0}\\
R_{2,1}=0:\quad \f{\al_{2,1}}{Q}\f{(\D_e-\D_{3,1})(\D_e-\D_{3,-1})}{\D_{3,1}\D_{3,-1}}=0\label{R21=0}
\end{align}
The necessary condition for consistency of these constraints is that one of the following relations is satisfied
\begin{align}
\D_{1,3}=\D_{3,1},\qquad \D_{1,3}=\D_{3,-1},\qquad \D_{1,-3}=\D_{3,1},\qquad \D_{1,-3}=\D_{3,-1}
\end{align}
Since $\Delta_{r,s}$ only depends on $b$ \eqref{b def} these equations are constraints on the central charge. Once one of them is fulfilled we can choose $\D_e$ equal to the coincident dimensions (say $\D_e=\D_{1,3}=\D_{3,1}$) and make the numerators of both $R_{1,2}(\D_e,c)$ and $R_{2,1}(\D_e,c)$ vanishing. There are six inequivalent solutions to these equations 
\begin{align}
&\al_{1,1}=0,\quad &&\D_e=\Delta_{1,3}=\Delta_{3,1}=1,&&\quad c=1 \label{al1}\\ 
&\al_{1,-1}=0,\quad &&\D_e=\Delta_{1,3}=\Delta_{3,1}=-3,&&\quad c=25\\
&\al_{1,-2}=0,\quad &&\D_e=\Delta_{1,3}=\Delta_{3,-1}=-2,&&\quad c=28\\
&\al_{2,1}=0,\quad &&\D_e=\Delta_{1,3}=\Delta_{3,-1}=3,&&\quad c=-2\\
&\al_{1,1}=0,\quad &&\D_e=\Delta_{1,-3}=\Delta_{3,-1}=4,&&\quad c=1\\
&\al_{1,-1}=0,\quad &&\D_e=\Delta_{1,-3}=\Delta_{3,-1}=0,&&\quad c=25 \label{al6}
\end{align}
Equation $\Delta_{r,s}=\Delta_{m,n}$ implies that either $\al_{r+n,s+m}=0$ or $\al_{r-n,s-m}=0$. We will often refer to these conditions on $\al$ as to solutions themselves, because they unambiguously fix the central charge while being more compact and structured than the particular values of $c$. These conditions are written in the first column. In the second column are listed values of the coincident dimensions, to which the external dimension must be equal. The last column depicts the corresponding central charge.\footnote{We have six instead of eight solutions here, because equations $\Delta_{1,3}=\Delta_{3,-1}$ and $\Delta_{1,-3}=\Delta_{3,1}$ can be converted to each other by the replacement $b\to b^{-1}$ exchanging $r\leftrightarrow s$ in $\D_{r,s}$ and using $\D_{r,s}=\D_{-r,-s}$. Since $b\to b^{-1}$ leaves the central charge invariant, both equations lead to the same conformal blocks.}

Solutions with $\D_e=0$ and $\D_e=1$ (the first and the last one) reduce the situation to the no-pole case and are of no interest here. The remaining four cases turn out to be quite different and are analyzed one by one. We emphasize in advance that conditions \eqref{al1}-\eqref{al6} only ensure vanishing of the numerators in $R_{1,2}$ and $R_{2,1}$. Alone, they are not sufficient to obtain a single-pole block .
\subsubsection*{\underline{Case: $\al_{1,-1}=0,\quad \D_e=\Delta_{1,3}=\Delta_{3,1}=-3,\quad c=25$}}
The parameters are adjusted so that the numerators of \eqref{R12=0}, \eqref{R21=0} vanish. However, for $\al_{1,-1}=0$ we also have $\D_{1,-3}=\D_{-3,1}=0$. Thus, for this specific value of the central charge the denominators in \eqref{R12=0}, \eqref{R21=0} also vanish and we have to resolve the ambiguity.
 
In other words, requesting that numerators of both $R_{1,2}$ and $R_{2,1}$ are zero fixes not only the external dimension, but also the central charge. The order in which we specify these values is important. If $\al_{1,-1}=0$ is set first, keeping $\D_e$ generic, the both terms are singular. On the other hand, first setting $\D_e=\D_{1,3}$ renders $R_{1,2}$ as well as any $R_{r\ge1,s\ge2}$ vanishing. However, $R_{2,1}(\D_{1,3},c)$ is not zero for generic central charge. The limit of this quantity as $\al_{1,-1}\to0$ is in fact finite 
\begin{equation}
\lim_{\al_{1,-1}\to0}R_{2,1}(\D_{1,3},c)=-18
\end{equation} 
Hence, the current set of parameters does not lead to a single-pole conformal block. In fact, it is even worse than that. Explicit computation gives
\begin{eqnarray}
H_\D(-3,25|q)=1+\frac{6 q}{\Delta }+\frac{36 (5+2 \Delta ) q^2}{\Delta  (5+4 \Delta )}+\frac{(840+96 \Delta ) q^3}{\Delta  (5+4 \Delta )}+\frac{24 \left(390+116 \Delta +7 \Delta ^2\right) q^4}{\Delta  (3+\Delta ) (5+4 \Delta )}+O\left(q^5\right) \label{H -3 25}
\end{eqnarray}
Additional pole at $\D=-5/4$ is expected due to non-vanishing $R_{2,1}$ which contributes a pole at $\D=\D_{2,1}$ ($=-5/4$ at $\al_{1,-1}=0$). The pole at $\D=-3$ however calls for an explanation. As turns out, condition $R_{r,s}=0$ alone it is not sufficient for the pole contribution at $\D=\D_{r,s}$ to be absent in \eqref{Zamolodchikov's elliptic}. The reason is that the coefficient $H_{\D_{r,-s}}$ might be singular so that the product $R_{r,s}H_{\D_{r,-s}}$ remains finite.  

Given this obstacle to handle, let us consider the situation more carefully. After we have set $\D_e=\D_{1,3}$ keeping the central charge generic, the general recurrence relation \eqref{Zamolodchikov's elliptic} reduces to
\begin{eqnarray}
H_{\D}(\D_{1,3},c|q)=1+\sum_{r\ge1}\f{R_{r,1}(\D_{1,3},c)}{\D-\D_{r,1}}q^rH_{\D_{r,-1}}(\D_{1,3},c|q)\label{s to 1}
\end{eqnarray}
The sum is restricted to $s=1$ because $R_{r,s}(\D_e,c)$ with $s\neq1$ vanish for $\D_e=\D_{1,3}$. Let us now substitute $\D=\D_{n,-1}$
\begin{eqnarray}
H_{\D_{n,-1}}(\D_{1,3},c|q)=1+\sum_{r\ge1}\f{R_{r,1}(\D_{1,3},c)}{\D_{n,-1}-\D_{r,1}}q^rH_{\D_{r,-1}}(\D_{1,3},c|q)
\end{eqnarray}
For some specific values of the central charge $\D_{n,-1}$ happen to coincide with $\D_{r,1}$. At $\al_{1,-1}=0$ this happens when $n=r+2$. Simplest example arise when $r=1$ and $n=3$ yielding $\Delta_{3,-1}=\Delta_{1,1}$. Hence, the $H_{\D_{3,-1}}(\D_{1,3},c|q)$ block is singular at $\al_{1,-1}=0$ so that the combination $R_{3,1}H_{\D_{3,-1}}$ is finite and contributes a pole at $\D=\D_{3,1}$ ($=-3$ at $\al_{1,-1}=0$) which precisely accounts for the extra pole in \eqref{H -3 25}. Moreover, it is possible to show that in fact an infinite amount of additional poles appears in higher orders by the same mechanism (see \ref{proof}). This should be contrasted with some cases to be discussed below, where only a finite amount of additional poles appear. 
\subsubsection*{\underline{Case: $\al_{1,-2}=0,\quad\D_e=\Delta_{1,3}=\Delta_{3,-1}=-2,\quad c=28$}}
Setting $\D_e=\Delta_{1,3}$ renders $R_{r\ge1,s\ge2}$ zero. Contrary to the $\al_{1,-1}=0$ case, $\al_{1,-2}=0$ does not lead to singularities in the denominators of \eqref{R12=0},\eqref{R21=0}. One can also show that any $R_{r\ge2,s=1}$ do vanish in the current setup, in agreement with the naive expectations. Hence, $R_{1,1}$ is the only non-zero coefficient. Nevertheless, we still do not obtain a single-pole conformal block. The reason again is that some conformal blocks entering residues in \eqref{Zamolodchikov's formula} are singular at this central charge. Indeed, consider for example $r=2, s=1$ contribution in \eqref{Zamolodchikov's elliptic}, where $H_{\D_{2,-1}}$ appears. We can show it to be singular by the same arguments as in the previous case. Formula \eqref{Zamolodchikov's elliptic} for $\D=\D_{2,-1}$ in the two lowest orders gives
\begin{eqnarray}
H_{\D_{2,-1}}(\D_e,c|q)=1+\f{R_{1,1}(\D_e,c)}{\D_{2,-1}-\D_{1,1}}qH_{\D_{1,-1}}(\D_e,c|q)+\dots
\end{eqnarray}
In the limit $\al_{1,-2}=0$ dimension $\Delta_{2,-1}=0$ and thus coincides with $\Delta_{1,1}$. 
Since $R_{1,1}$ is non-zero, we find that the conformal block $H_{\Delta_{2,-1}}$ is singular. For generic $\D$ in recursion \eqref{s to 1} this singularity cancels the zero in $R_{2,1}$ leading to a non-vanishing contribution of the term with pole at $\D=\D_{2,1}=-2$. So this is still not a sought-for single-pole block. More importantly, just as in the previous example this is only the tip of the iceberg: infinitely many different poles are in fact present in the case at hand. As we argue in subsection \ref{general situation sec}, this kind of behavior is always met at $c>1$ while the true finite-pole blocks only appear at $c\le1$.

\subsubsection*{\underline{Case: $\al_{2,1}=0,\quad \D_e=\Delta_{1,3}=\Delta_{3,-1}=3,\quad c=-2$}}
This case is similar to the first one we have considered (with $\al_{1,-1}=0$) in that there is a zero in the denominator of $R_{2,1}$ ($\D_{3,1}=0$) upon substitution of $\al_{2,1}=0$. Unlike the previous case, the additional factor $\al_{2,1}$ in $R_{2,1}$ \eqref{R21=0} vanish, too. So overall $R_{2,1}=0$. $R_{3,1}$ is, however, finite
\begin{equation}
R_{3,1}(\D_e,c)= \f{\al_{3,1}}{\al_{2,1}}\f{(\D_e-\D_{5,1})(\D_e-\D_{5,-1})}{\D_{5,1}\D_{5,-1}}R_{2,1}(\D_e,c),\qquad \lim_{c\to-2}R_{3,1}(\D_{1,3},c)=-15 \label{R31}
\end{equation}
The reason is that $\al_{3,1}$ is non-vanishing with the current parameters so no additional zero arise in the numerator to compensate for vanishing $\D_{3,1}$. Nevertheless, any $R_{r,1}$ with $r\ge4$ contains $(\D_e-\D_{7,-1})$ factor which is vanishing since $\D_{7,-1}=\D_e=3$ at $\al_{2,1}=0$. We are led to conclude that only $R_{1,1}$ and $R_{3,1}$ are non-zero. They key feature of the current case is that $R_{3,1}$ contribution does not add a new pole since $\D_{3,1}=\D_{1,1}=0$ at $\al_{2,1}=0$.

As we have seen in the previous examples vanishing of $R_{r,s}$ is not enough for the pole at $\D=\D_{r,s}$ to be absent in conformal block. The accompanying factor $H_{\D_{r,-s}}$ must be non-singular. Recall that upon setting $\D_e=\D_{1,3}$ the general recursion is reduced to \eqref{s to 1}. Hence, if $\D_e=\D_{1,3}$ we only have to worry about $B_{\D_{n\ge1,-1}}(\D_{1,3},c|q)$ being singular at $\al_{2,1}=0$. A singularity in $B_{\D_{n,-1}}(\D_{1,3,},c|q)$ may only arise if (1) $\D_{n,-1}=\D_{r,1}$ for some $r$ and (2) the coefficient $R_{r,1}(\D_{1,3},c)$ is non-zero for that $r$. Since the only non-zero $R_{r,1}$ are found for $r=1$ and $r=3$ and $\D_{1,1}=\D_{3,1}=0$ condition (1) requires $\D_{n,-1}=0$ which is never true for $\al_{2,1}=0$. Therefore, none of the conformal blocks entering reduced recursion \eqref{s to 1} is singular at $\al_{2,1}=0$. Finally, we have obtained a true one-pole conformal block which satisfies 
\begin{equation}
H_{\D}(\D_{1,3},c|q)=1+\f{R_{1,1}(\D_{1,3},c)}{\D-\D_{1,1}}qH_{\D_{1,-1}}(\D_{1,3},c|q)+\f{R_{3,1}(\D_{1,3},c)}{\D-\D_{3,1}}q^3H_{\D_{3,-1}}(\D_{1,3},c|q) \label{1 pole implicit}
\end{equation} 
This equation could of course be simplified by substituting explicit values for $\D_{1,3},\D_{1,1}$ and $c$ but we find the current form more instructive.

It may seem that equation \eqref{1 pole implicit} allows to easily find the full $q$-dependence of the conformal block at hand. Indeed, setting $\D=\D_{1,-1}$ and $\D=\D_{3,-1}$ we obtain a pair of equations which can be considered as a linear system for $H_{\D_{1,-1}}$ and $H_{\D_{3,-1}}$. Solving for these variables and substituting the result back to \eqref{1 pole implicit} would amount to finding $H_{\D}$ explicitly, to all orders in $q$.

The caveat is that equation \eqref{1 pole implicit} does not hold for some specific values of $\D$, in particular for $\D=\D_{1,-1}$. The problem again roots in the order of limits. 
Let us set $\D=\D_{1,-1}$ in \eqref{s to 1}
\begin{eqnarray}
H_{\D_{1,-1}}(\D_{1,3},c|q)=1+\sum_{r\ge1}\f{R_{r,1}(\D_{1,3},c)}{\D_{1,-1}-\D_{r,1}}q^rH_{\D_{r,-1}}(\D_{1,3},c|q)
\end{eqnarray}
The limit of this recursion as $\al_{2,1}=0$ is different than for a generic $\D$. Despite conformal blocks $H_{\D_{r,-1}}$ are not singular at $\al_{2,1}=0$ the denominator $\D_{n,1}-\D_{r,-1}$ can vanish and rescue the $R_{r,1}$ term. When $\al_{2,1}=0$ we have $\D_{n,1}=\D_{r,-1}$ when $r=n+4$. In particular, $\D_{5,1}=\D_{1,-1}$. In fact 
\begin{eqnarray}
\lim_{c\to-2}\f{R_{5,1}(\D_{1,3},c)}{\D_{1,-1}-\D_{5,1}}=14
\end{eqnarray}
Hence, for $\D=\D_{1,-1}$ an additional term appears in the relation \eqref{1 pole implicit}
\begin{multline}
H_{\D_{1,-1}}(\D_{1,3},c|q)=1+\f{R_{1,1}(\D_{1,3},c)}{\D-\D_{1,1}}qH_{\D_{1,-1}}(\D_{1,3},c|q)+\\\f{R_{3,1}(\D_{1,3},c)}{\D-\D_{3,1}}q^3H_{\D_{3,-1}}(\D_{1,3},c|q)+14q^5H_{\D_{5,-1}}(\D_{1,3},c)
\end{multline} 
In turn, the counterpart of \eqref{1 pole implicit} valid at $\D=\D_{5,-1}$ contains  $H_{\D_{9,-1}}$ and so forth. As a consequence, equation \eqref{1 pole implicit} can not be reduced to a finite linear system on a subset of $H_{\Delta_{n,-1}}$.

Nevertheless it turns out, somewhat surprisingly, that after we have a set of parameters for which the amount of $\D$-poles is finite, complete and explicit expressions for conformal blocks are achievable. It is known (see for example \cite{Billo:2013fi}) that toric conformal blocks can be expressed in terms of the modular functions $E_2, E_4,E_6$. However, in general arbitrary powers of these modular functions are present. Following \cite{Beccaria:2016nnb} we observe that the finite-pole cases under discussion feature another remarkable property: they only contain modular forms up to a finite weight which seems to be proportional to the number of poles. 

As explained, formula \eqref{1 pole implicit} only ensures that there is a single pole in conformal block at $\D=0$, but does not simplify determination of the full $q$-dependence. On the other hand, coefficients in conformal block $q$-expansion can be computed to any finite order with the general recurrence formula \eqref{coeff recursion} or directly from the definition \eqref{CB trace def}. Specifying these generic expressions to $\D_e=3, c=-2$ we find
\begin{eqnarray}
H_\D(3,-2|q)=1+\frac{3 q}{\Delta }+\frac{9 q^2}{\Delta }+\frac{12 q^3}{\Delta \
}+\frac{21 q^4}{\Delta }+\frac{18 q^5}{\Delta }+O\left(q^6\right)
\end{eqnarray}
In this expansion one recognizes the second Eisenstein series\footnote{We use $E_2(q)=1-24\sum_{n=1}^\infty\f{nq^n}{1-q^n}$, $E_4(q)=1+240\sum_{n=1}^\infty\f{n^3q^n}{1-q^n}$.}
\begin{eqnarray}
\boxed{H_\D(3,-2|q)=1+\f{1-E_2(q)}{8\D}} \label{single pole explicit}
\end{eqnarray}
This conjectural relation can be verified to any desired order. Unfortunately, due to obstacles outlined above, there seems to be no easy way to either derive this formula or test it to all orders using Zamolodchikov's relation. In this regard our analysis barely adds something new to results of \cite{Beccaria:2016nnb}.
\subsubsection*{\underline{Case: $\al_{1,1}=0,\quad \D_e=\Delta_{1,-3}=\Delta_{3,-1}=4,\quad c=1$}}
This case shows yet another interesting feature but will not be considered in the same detail as the preceding ones. Similarly to the previous ($\al_{2,1}=0$) case fixing the external dimension $\D_e=\D_{1,-3}$ and taking the $\al_{1,1}\to0$ limit leaves only two of the $R_{r,s}$ coefficients non-zero, namely $R_{1,1}$ and $R_{2,1}$. Further, none of the coefficients $H_{\D_{r,-s}}$ relevant for the recursion in this case are singular. However, $\D_{2,1}=1/4$ at $c=1$ and does not coincide with $\D_{1,1}=0$. Hence we do not obtain a single-pole conformal block. Nevertheless, in contrast to the cases with $c=25$ and $c=28$ this is the only additional pole appearing and the current set of parameters provides a true two-pole conformal block. Omitting the computation details, we only present the final result
\begin{eqnarray}
\boxed{H_\D(4,1|q)=1+\f{1-E_2(q)}{4\D-1}+\f{E_2(q)^2-E_4(q)}{48\D(4\D-1)}} \label{two pole explicit}
\end{eqnarray}
Again, this formula can be verified against the general conformal block expansion to any order in $q$.

\subsection{General situation} \label{general situation sec}
Our study of what was supposed to be a single-pole conformal block was not as plain as one could expect given very explicit formula \eqref{Zamolodchikov's elliptic} on the disposal. We have seen that some of the candidate relations \eqref{al1}-\eqref{al6} instead lead to zero-pole or two-pole blocks, while some others do not give finite-pole solutions at all. Nevertheless, the strategy we have chosen seems reasonable enough to generalize.

To this end, one can pick two numbers $N,M$ and require that $R_{1,N+1}$ and $R_{M+1,1}$ are zero. Explicitly, this reads
\begin{align}
&R_{1,M+1}=0:\quad &&\f{\al_{1,M}}{\al_{1,M-1}}\f{(\D_e-\D_{1,2M+1})(\D_e-\D_{1,-2M-1})}{\D_{1,2M+1}\D_{1,-2M-1}}R_{1,M}(\D_e,c)=0\\
&R_{N+1,1}=0:\quad &&\f{\al_{N,1}}{\al_{N-1,1}}\f{(\D_e-\D_{2N+1,1})(\D_e-\D_{2N+1,-1})}{\D_{2N+1,1}\D_{2N+1,-1}}R_{N,1}(\D_e,c)=0
\end{align}
If $R_{1,M}(\D_e,c)=0$ or $R_{N,1}(\D_e,c)=0$ the situation reduces to the previous step and does not have to be considered. Remaining possibilities are described by equations 
\begin{align}
&\D_{1,2M+1}=\D_{2N+1,1},&\quad &\D_{1,2M+1}=\D_{2N+1,-1},&\quad &\D_{1,-2M-1}=\D_{2N+1,1},&\quad &\D_{1,-2M-1}=\D_{2N+1,-1}
\nn\\[8pt]
&\al_{N,-M}=0,&&\al_{N,-M-1}=0,&&\al_{N+1,-M}=0,&&\al_{N+1,-M-1}=0\nn\\
&\al_{N+1, M+1}=0,&&\al_{N+1, M}=0,&&\al_{N, M+1}=0,&&\al_{N, M}=0 \label{general cond}
\end{align}
The first line here presents original conditions on the dimensions. Each of these constraints has two solutions of the form $\al_{n,m}=0$ for some $n,m$, similarly to those found in \eqref{al1}-\eqref{al6}. These are written in the remaining two lines. At the example of $N=M=1$ we saw that some of these solutions do not lead to the finite-pole blocks. By direct computer-assisted computations for several $N,M$ up to and including $N\times M=6$ we observe a general trend: only those solutions $\al_{n,m}=0$ for which $n$ and $m$ are of the same sign (given in the last line of \eqref{general cond}) lead to the finite-pole blocks. When $\al_{n,m}=0$ the central charge is
\begin{eqnarray}
c=1-6\f{(n-m)^2}{nm} \label{minimal models c}
\end{eqnarray}
If $n,m$ are of the same sign this formula precisely describes the central charges of the minimal models (including $c=1$). Thus, we conjecture that the finite-pole blocks only exist in theories with the central charges equal to those of the minimal models. We prove this conjecture for $M=1$, $N$ arbitrary in appendix \ref{proof}. In the remainder of the text we assume the conjecture to be true. 

For a given value of the central charge there is more then one external dimension leading to a finite-pole block. Indeed, from \eqref{general cond} we read off four external dimensions allowed at $\al_{N,M}=0$
\begin{eqnarray}
\al_{N,M}=0:\qquad \D_e\in \{\D_{1,2M-1}, \D_{1,2M+1}, \D_{1,-2M+1},\D_{1,-2M-1}\}
\end{eqnarray}
Explicitly these are
\begin{eqnarray}
\al_{N,M}=0:\qquad \D_e\in \{(N-1)(M-1),\, NM+N-M,\, NM-N+M,\,(N+1)(M+1)\} \label{ext dim expl}
\end{eqnarray}
Now, note that conditions $\al_{N,M}=0$ and $\al_{N',M'}=0$ lead to the same central charge if $N'/N=M'/M$. For a given central charge \eqref{minimal models c} assume that $n, m$ are coprime. Then, the other possibilities leading to this central charge are described by
$\al_{pN,pM}=0$ for any $p\in\mathbb{Z}\setminus \{0\}$. However, the external dimensions \eqref{ext dim expl} are not invariant under the rescaling $n,m\to pn,pm$. Hence, for central charge \eqref{minimal models c} the following external dimensions are possible\footnote{Note that it is sufficient to rescale only two dimensions from \eqref{ext dim expl} since the remaining two are obtained by reflection $N,M\to-N,-M$ which is accounted for when $p$ takes negative values.}
\begin{eqnarray}
\D_e\in\bigcup\limits_{\substack{n,m\in\mathbb{Z}_+\\p\in\mathbb{Z}\setminus \{0\}}}\Big\{(pn-1)(pm-1),\, p^2nm+pn-pm\Big\} \label{ext dimensions}
\end{eqnarray}
Thus, an infinite number of external dimensions provide a finite-pole truncation at a given central charge. All the examples found in \cite{Beccaria:2016nnb} fit within this classification. However, the exact number of poles for a particular solution from \eqref{ext dimensions} seems quite irregular as a function of $n, m, p$. This issue receives a surprisingly simple solution in subsection \ref{mk two poles toric sec} where the modular transformations are considered.

One more important note is in order. In \cite{Beccaria:2016nnb} there is a second instance of a single-pole conformal block, different from \eqref{single pole explicit}. It appears when $\D_e=2,c=0$. In our considerations it arises as a solution to $\al_{N+1,M+1}=0, \D_e=\D_{1, 2M+1}$ for $N=2, M=1$. Hence, this new single-pole block is found where one naively expects a two-pole block. In this case, the reason can be traced back to the fact that the poles accompanying the only two non-vanishing coefficients $R_{1,1}$ and $R_{2,1}$ coincide at $c=0$: $\D_{1,1}=\D_{2,1}=0$. This is the possibility we have mentioned in passing. Solutions \eqref{general cond}, although not initially intended to account for such cases, seem to also handle them. It is very plausible that the conformal blocks with the external dimensions \eqref{ext dimensions} exhaust all finite-pole blocks at a given minimal model central charge \eqref{minimal models c}.
\section{Exact spheric conformal blocks}
We now proceed to the discussion of the finite-pole spheric conformal blocks which share most of the qualitative features with the toric ones. Four-point spheric conformal block $B_\D(\D_i,c|x)$ depends on seven parameters in total: conformal cross-ratio $x$, one internal $\D$ and four external dimensions $\D_i$ $( i=1,2,3,4$), and the central charge $c$. We will not present the definition in terms of the correlation functions or the Virasoro algebra but simply state the recurrence equation instead. In order to do this, introduce the elliptic spheric block $H_\D(\D_i|q)$ as 
\begin{eqnarray}
B_{\D}(\D_i,c|x)=F_{\D}(\D_i,c|x)H_\D(\D_i,c|q)\label{spheric cb}
\end{eqnarray}
with
\begin{eqnarray}
F_{\D}(\D_i,c|x)=(16q)^{\D-\f{c-1}{24}}x^{\f{c-1}{24}-\D_1-\D_2}(1-x)^{\f{c-1}{24}-\D_2-\D_3}\theta_3(q)^{\f{c-1}{2}-4(\D_1+\D_2+\D_3+\D_4)} \label{spheric prefactor}
\end{eqnarray}
Here\footnote{Note that we use the same notations for analogous quantities in the toric and the spheric case, for example the nome $q$, the blocks $B, H$, the coefficients $R_{r,s}$ below etc. The meaning should be clear from the context.}  
\begin{eqnarray}
q=e^{i\pi\mathcal{T}},\qquad \mathcal{T}=i\f{K(1-x)}{K(x)} \label{spheric nome}
\end{eqnarray}
For our conventions on elliptic functions see appendix \ref{special functions}. The elliptic block satisfies recursion of the type \eqref{Zamolodchikov's elliptic} 
\begin{eqnarray}
H_{\D}(\D_i,c|q)=1+\sum_{r,s\ge1}\f{R_{r,s}(\D_i,c)}{\D-\D_{r,s}}(16q)^{rs}H_{\D_{r,-s}}(\D_i, c|q)
\end{eqnarray}
The residues are given by $R_{r,s}(\D_i,c)=A_{r,s}(c)P_{r,s}(\D_i,c)$ where the part independent of $\D_i$ is the same as in the toric case
\begin{eqnarray}
A_{r,s}=\f{\al_{r,s}}{Q}\prod_{n=0}^{r-1}\prod_{m=0}^{s-1}\f1{\D'_{2n+1,2m+1}\D_{2n+1,-2m-1}}\label{A}
\end{eqnarray}
while the remainder is given by
\begin{eqnarray}
P_{r,s}(\D_i,c)=\prod_{i=1}^{4}\,\,\prod_{\substack{n=1-r\\\Delta n=2}}^{r-1}\,\,\prod^{s-1}_{\substack{m=1-s\\\Delta m=2}}(\mu_i-\al_{n,m})\label{P}
\end{eqnarray}
Here $\mu_i$ are related to $\alpha_i$ as
\begin{eqnarray}
\mu_1=\al_1+\al_2,\qquad \mu_2=\al_1-\al_2,\qquad \mu_3=\al_3+\al_4,\qquad \mu_4=-\al_3+\al_4 \label{mu to al}
\end{eqnarray}
Notice a slight notation abuse here since $R_{r,s}(\D_i,c)$ are functions of $\al_i$ rather then $\D_i$.

The backbone for our analysis of the toric case was provided by the fact that the toric residues $R_{r,s}(\Delta_e, c)$ \eqref{residue} have a nested structure of zeros, i.e. that vanishing of $R_{n,m}$ also implies vanishing of $R_{r\ge n,s\ge m}$. The situation in the spheric case is analogous but somewhat different. Most importantly, in the spheric residues $R_{r,s}$ \eqref{P} the products over $n$ and $m$ are taken with step 2, i.e. they range over
\begin{eqnarray}
n=1-r,3-r,\dots, r-3, r-1,\qquad m=1-s,3-s,\dots, s-3, s-1
\end{eqnarray}
Hence, the zeros of $R_{n,m}$ are not in general inherited by all $R_{r\ge n,s\ge m}$ but only by those that have $r=n+2k, s=m+2l$ for  $k,l\in \mathbb{Z}_+$. It is therefore useful to separate pairs $r,s$ into four families
\begin{eqnarray}
\bigcup\limits_{r,s\ge1}R_{r,s}=\bigcup\limits_{n,m\ge1}\left\{R_{2n-1,2m-1}, R_{2n-1,2m}, R_{2n,2m-1}, R_{2n,2m}\right\} \label{residue families}
\end{eqnarray}
so that within each family the zeros are inherited by the higher order terms from the smaller ones.

Thus, in the spheric case we have four independent residue grids and four external dimensions available for adjusting, so the net situation is quite similar to the toric case. Making all but the finite amount of residues to vanish within each family implies fine-tuning of the central charge. Thus, finite-pole blocks are typically encountered when the central charge and all the external dimensions are set to specific values. The exception is again the pole-free case to be discussed right below.
\subsection{No poles}
To obtain a zero-pole conformal block the seeds of all the four residue families \eqref{residue families} have to vanish
\begin{align}
R_{1,1}=0:\qquad & \prod_{i=1}^4 (\mu_i-\al_{0,0})=0 \label{p11}\\
R_{1,2}=0:\qquad & \prod_{i=1}^4 (\mu_i-\al_{0,1})(\mu_i+\al_{0,1})=0 \label{p12}\\
R_{2,1}=0:\qquad & \prod_{i=1}^4 (\mu_i-\al_{1,0})(\mu_i+\al_{1,0})=0 \label{p21}\\
R_{2,2}=0:\qquad & \prod_{i=1}^4 (\mu_i-\al_{1,1})(\mu_i+\al_{1,1})(\mu_i-\al_{1,-1})(\mu_i+\al_{1,-1})=0 \label{p22}
\end{align}
Relation $\al_{-n,-m}=-\al_{n,m}$ was used here. We assume that the central charge is generic so that we do not have to consider potentially singular multipliers \eqref{A}. This also implies that no roots in the above equations coincide.

Equation \eqref{p11} fixes one of the parameters uniquely. Modulo permutation of dimensions we can assume that it fixes $\mu_1$ and choose $\mu_1=\al_{0,0}$. Let us view the next equation \eqref{p12} as a constraint on $\mu_2$. Then we put $\mu_2=\al_{0,1}$. The alternative choice ($\mu_2=-\al_{0,1}$) leads to the swap of dimensions $\al_1\leftrightarrow\al_2$ \eqref{mu to al}. Similarly, we set $\mu_3=\al_{1,0}$ to fulfill \eqref{p21}. Finally, equation \eqref{p22} provide four options to pick from for $\mu_4$: $\mu_4=\pm\al_{1,1}$ or $\mu_4=\pm\al_{1,-1}$. The sign choice in both cases is again related to a swap of dimensions ($\al_3\leftrightarrow\al_4$) that we will not take into account. Thus, finally, we obtain two inequivalent solutions (those which are not related by a permutation of external dimensions) which differ by the choice of $\mu_4$
\begin{align}
&\mu_1 = \al_{0,0},\quad \mu_2 = \al_{0,1},\quad \mu_3 = \al_{1,0},\quad \mu_4 = \al_{1,1}\\
&\mu_1 = \al_{0,0},\quad \mu_2 = \al_{0,1},\quad \mu_3 = \al_{1,0},\quad \mu_4 = \al_{1,-1}
\end{align}
In terms of the central charge parameter $b$ the external dimensions for these solutions are
\begin{align}
&\D_1=\D_2=\D_3=\f{8+3b^{-2}+4 b^2}{16},\quad\D_4=\f{4+3b^{-2}}{16} \label{1 no pole family}\\
&\D_1=\D_2=\D_3=\f{8+3b^{-2}+4 b^2}{16},\quad\D_4=\f{3(4+b^{-2})}{16} \label{2 no pole family}
\end{align}
In case of $c=1$ ($b=i$) the first solution reduces to $\D_1=\dots=\D_4=1/16$, the Ashkin-Teller model, while the second solution reduces to $\D_1=\dots=\D_4=15/16$ at $c=25$ ($b=1$), a related CFT. To our knowledge, these two cases are the only examples of the closed-form four-point blocks with continuous intermediate dimension available in the literature. As our analysis shows each of these models in fact belongs to a one-parametric family extending to an arbitrary central charge. In all these cases the elliptic conformal block is trivial
\begin{eqnarray}
H_\D(\D_i,c|q)=1
\end{eqnarray}
As a side note we point out that these two families are reminiscent of the two possible values of $\D_e$ featuring in the toric no-pole case. This analogy strengthens further when the modular transformations are considered, see subsection \ref{sec mk no pole toric}.
\subsection{Single pole}
It is not our aim in the present subsection to describe all the single-pole spheric blocks. Instead we will discuss one simplest example. To this end, let us lighten $R_{1,1}=0$ constraint of the no-pole situation and replace it with $R_{1,3}=R_{3,1}=0$. This will only affect the choice of $\mu_1$ which now has to satisfy two conditions simultaneously
\begin{eqnarray}
R_{1,3}=0:\quad (\mu_1-\al_{0,2})(\mu_1+\al_{0,2})=0,\quad R_{3,1}=0:\quad (\mu_1-\al_{2,0})(\mu_1+\al_{2,0})=0
\end{eqnarray}
They are not compatible for a generic central charge. But they are compatible for, say, $\al_{2,0}=-\al_{0,2}$ ($b=i, c=1$) in which case we can choose $\mu_1=\al_{2,0}$. Out of two possibilities for $\mu_4=\al_{1,1}, \mu_4=\al_{1,-1}$ we pick the former. Hence
\begin{align}
\mu_1 = \al_{2,0},\quad \mu_2 = \al_{0,1},\quad \mu_3 = \al_{1,0},\quad \mu_4 = \al_{1,1},\qquad c=1
\end{align}
The corresponding dimensions are
\begin{eqnarray}
\D_1=\f{1}{16},\quad\D_2=\f{9}{16},\quad\D_3=\f{1}{16},\quad\D_4=\f{9}{16}
\end{eqnarray}
We shall denote these dimensions collectively by $\bar{\D}_i$. The first several terms of the elliptic block expansion in this case are
\begin{eqnarray}
H_\D(\bar{\D}_i,1|q)=1+2\frac{-q+2 q^2-4 q^3+4 q^4-6 q^5+8 q^6-8q^7+8 q^8-13 q^9+12 q^{10}}{\D}+O(q^{11})
\end{eqnarray}
from which the closed form expression is easily inferred
\begin{eqnarray}
\boxed{H_\D(\bar{\D}_i,1|q)=1-\f{E_2(q)+\theta_2^4(q)-\theta_4^4(q)}{12\D} \label{1 pole spheric}}
\end{eqnarray}
This example illustrates that there are non-trivial finite-pole spheric conformal blocks. Due to increased number of parameters producing more examples, let alone obtaining the full classification of these finite-pole cases seems much harder than in the toric case and we do not attempt it in the present paper. However, the generic features seem to be common to both cases.
\section{Modular transformations}
Having closed-form expressions for the conformal blocks invites to study their modular properties. Let us mention that in paper \cite{Beccaria:2016nnb} the finite-pole blocks were tested against the modular anomaly equation \cite{Billo:2014bja} to find a complete agreement. Our aim is to describe the modular transformations in an explicit way and construct the corresponding kernels. 
\subsection{Toric}
Toric conformal block depends on the nome $q=e^{2\pi i\tau}$. The modular group is generated by the transformations $T:\tau\to\tau+1$ and $S:\tau\to-1/\tau$. $T$ acts trivially (even on  a generic conformal block) so we will only be concerned with the action of $S$ to which we refer simply as to the modular transformation.

Transformation properties of the toric one-point function imply the following transformation law of conformal blocks
\begin{eqnarray}
B_{\D}(\D_e,c|\widetilde{q}\,)=\tau^{\D_e} \int\limits_{CFT spectrum}{d\D}\, M_{\D\D'}(\D_e,c)B_{\D'}(\D_e,c|q) \label{modular transform toric}
\end{eqnarray}
Here $\widetilde{q}=q^{-2\pi i/\tau}$ and $M_{\D\D'}(\D_e,c)$ is the $q$-independent kernel of this linear transformation called the modular kernel. An important concern is what is the domain of integration in \eqref{modular transform toric}. Conformal symmetry alone does not answer this question. There may be more than one consistent choice. We assume that the spectrum is continuous and given by 
\begin{eqnarray}
\Delta\in \f{c-1}{24}+\mathbb{R}_+ \label{continious spectrum}
\end{eqnarray}
which is very natural from the standpoint of the Liouville theory and its generalizations, see \cite{Ribault:2015sxa}. Moreover, this choice will prove to be consistent.

It is convenient to introduce the following (slightly asymmetric) Liouville-type parametrization for the dimensions
\begin{eqnarray}
\D=Q^2/4-\al^2,\qquad \D'=Q^2/4-\al'^2,\qquad \D_e=\mu(Q-\mu) \label{dimension parametrization}
\end{eqnarray} 
Note that spectrum \eqref{continious spectrum} corresponds to $\al\in i\mathbb{R}_+$. In terms of the new variables equation \eqref{modular transform toric} becomes\footnote{In the following the modular kernel and conformal blocks are represented as either functions of the dimensions $\D,\D',\D_e$ or momentums $\al,\al',\mu$. To keep the notation lightweight the representations are sometimes mixed together, as in equation \eqref{modular transform toric momentums}. Hopefully, this will not lead to a confusion.}
\begin{eqnarray}
B_{\D}(\D_e,c|\widetilde{q}\,)=\tau^{\D_e} \int\limits_{i\mathbb{R}_+}{d\al'}\, M_{\al\al'}(\mu,b)B_{\D'}(\D_e,c|q) \label{modular transform toric momentums}
\end{eqnarray} 

For generic irrational $b$ the modular kernel is known in the form of integral \cite{PT3} or series \cite{Nemkov:2015zha} representation. However, these results do not apply to $c<1$ and hence to any of the finite-pole blocks (except the no-pole case). Nevertheless, in \cite{GMMnonpert, Nemkov:2015zha} a set of equations valid for arbitrary values of $c$ was derived and we will make use of these equations to infer the modular kernel. To cast the equations in a simple form we introduce the following renormalization of conformal block
\begin{eqnarray}
B_{\D'}(\D_e,c|q)=V_{\al}(\mu, b)\,\, \mathcal{B}_{\D'}(\D_e,c|q) \label{cb renormalization}
\end{eqnarray}
which leads to the following renormalization of the modular kernel
\begin{eqnarray}
M_{\al\al'}(\mu,b)=N_{\al\al'}(\mu, b)\,\,\mathcal{M}_{\al\al'}(\mu,b) \label{mk renormalization}
\end{eqnarray}
Here 
\begin{eqnarray}
N_{\al\al'}(\mu, b)=\f{V_{\al}(\mu,b)}{V_{\al'}(\mu,b)},\qquad V_{\al}(\mu,b)=\f{\Gamma_b(Q+2\al)\Gamma_b(Q-2\al)}{\Gamma_b(Q-\mu+2\al)\Gamma_b(Q-\mu-2\al)} \label{toric vertex}
\end{eqnarray}
and $\Gamma_b(z)$ is the double Gamma function defined in appendix \ref{special functions}. In terms of $\mathcal{M}_{\al\al'}(\mu,b)$ the equations are rather simple. The first one reads
\begin{eqnarray}
\left(\f{\sin\pi b(2\al+\mu)}{\sin{2\pi b\al}} e^{\f{b}2\p_\al}+\f{\sin\pi b(2\al-\mu)}{\sin{2\pi b\al}} e^{-\f{b}2\p_\al}\right)\mathcal{M}_{\al\al'}(\mu,b)=2\cos{2\pi b\al'}\mathcal{M}_{\al\al'}(\mu,b) \label{mk equation}
\end{eqnarray}
There are two more equations involving shift operators in $\al'$ and $\mu$, but we do not have to consider them. The equation with the shifts in $\al'$ follows from \eqref{mk equation} and the condition that the modular transform squares to unity
\begin{eqnarray}
\int_{i\mathbb{R}_+}d\al' M_{\al\al'}M_{\al'\al''}=\delta(\al-\al'') \label{mk unit square}
\end{eqnarray} 
which will be simpler to impose by hands. The equation with the shifts in $\mu$ is not relevant since we will fix the value of $\mu$ (corresponding to the external dimension) beforehand.

Two more remarks are in order. Equation \eqref{mk equation} is linear of second order and hence the solution space is two-dimensional. The proper choice is to pick the even function of $\al$ since the original modular kernel $M_{\D\D'}$ depends only on $\D$ which is an even function of $\al$ \eqref{dimension parametrization}. Next, note that the equation involves the shifts with values $b/2$ the solution is determined up to a periodic in $\al$ function with the period $b/2$. In fact, since the modular kernel must only depend on $c$ and not on $b$ separately the symmetry $b\to b^{-1}$ must be manifest in $\mathcal{M}_{\al\al'}(\mu,b)$. This further reduces the ambiguity up to a multiplier which is both $b/2$- and $b^{-1}/2$-periodic. For generic $b$ this fixes the solution uniquely \cite{Nemkov:2015zha}. For the finite-pole blocks however $b^2$ is rational and $b/2$-, $b^{-1}/2$-periodic function is not necessarily a constant. Still, we will be able to guess this remaining multiplier.

\subsubsection{No poles} \label{sec mk no pole toric}
Let us see how all this works for the no-pole conformal blocks, which are found for generic central charge at $\D_e=0$ or $\D_e=1$. First, let us choose $\mu=0$ which realizes the $\D_e=0$ scenario. The general equation \eqref{mk equation} then reads
\begin{eqnarray}
\left(e^{\f{b}2\p_\al}+e^{-\f{b}2\p_\al}\right)\mathcal{M}_{\al\al'}(0,b)=2\cos{2\pi b\al'}\mathcal{M}_{\al\al'}(0,b)
\end{eqnarray}
A possible solution is given by
\begin{eqnarray}
\boxed{\mathcal{M}_{\al\al'}(0,b)=2\s{2}\cos{4\pi \al\al'}\label{Fourier kernel}}
\end{eqnarray}
and we have introduced the normalization factor to satisfy \eqref{mk unit square}. Note that the renormalization $N_{\al\al'}$ \eqref{toric vertex} becomes trivial at $\mu=0$ hence \eqref{Fourier kernel} is the complete answer in this case. Indeed, the toric block with $\D_e=0$ reduces to the Virasoro character \eqref{character} which is well known to transform with the Fourier kernel
\begin{eqnarray}
\chi_\D(c|\widetilde{q}\,)=2\s{2}\int_{i\mathbb{R}_+}d\al'\, \cos{4\pi \al\al'}\chi_{\D'}(c|q)=\s{2}\int_{i\mathbb{R}}d\al'\, e^{4\pi i\al\al'}\chi_{\D'}(c|q)
\end{eqnarray}
This a simple exercise in the gaussian integration where the modular properties of the Dedekind eta function must be taken into account \eqref{special functions}. Note that in \eqref{Fourier kernel} we have implicitly chosen an undetermined periodic multiplier. Further we will clarify how this choice is made in general.

We now turn to the $\D_e=1$ case which appears to be somewhat different, despite the conformal block being exactly the same. Due to technical obstructions for a generic $c$, we will temporarily limit our attention to $b=i$ ($c=1$). The final answer will be valid for any $c$. For $b=i$ the choice $\mu=b=i$ provides $\D_e=1$. With this set of parameters equation \eqref{mk equation} becomes
\begin{eqnarray}
\left(e^{\f{i}{2}\p_\al}+e^{-\f{i}{2}\p_\al}\right)\mathcal{M_{\al\al'}}(i,i)=-2\cos{2\pi \al'}\mathcal{M_{\al\al'}}(i,i)
\end{eqnarray}
A possible solution is
\begin{eqnarray}
\mathcal{M_{\al\al'}}(i,i)=2\s{2}\,\sin{4\pi\al\al'}\f{\sin{2\pi i\al}}{\sin{2\pi i\al'}}
\end{eqnarray}
while the normalization factor reduces to
\begin{eqnarray}
N_{\al\al'}(i,i)=\f{\al'}{\al}\f{\sin{2\pi i\al}}{\sin{2\pi i\al'}}
\end{eqnarray}
so that the original modular matrix reads
\begin{eqnarray}
M_{\al\al'}(i,i)=2\s{2}\,\f{\al'}{\al}\sin{4\pi\al\al'}\times\left(\f{\sin{2\pi i\al}}{\sin{2\pi i\al'}}\right)^2
\end{eqnarray}
Note that thus obtained solution is only defined up to a $b/2$-, $b^{-1}/2$-periodic factor which for $b=i$ means that it is $i/2$-periodic. In the above formula the factor in braces is $i/2$-periodic and hence it is not demanded by the structure of equations. As turns out, omitting it gives the right modular kernel. Moreover, this kernel is valid for any value of $b$ if $\mu$ is such that $\D_e=1$. Hence we write
\begin{eqnarray}
\boxed{M_{\al\al'}(\bar{\mu},b)=2\s{2}\,\f{\al'}{\al}\sin{4\pi\al\al'}} \label{no pole sine kernel}
\end{eqnarray}
here $\bar{\mu}$ is any of solutions to $\mu(Q-\mu)=\D_e=1$. Indeed, 
\begin{multline}
\int_{i\mathbb{R}_+}d\al' M_{\al\al'}(\bar{\mu}, b)B_{\D'}(1,c|q)=
\f{\s{2}}{\al}\int_{i\mathbb{R}}d\al'\,\al'e^{4\pi i\al\al'}\chi_{\D'}(1|q)\\=\f{\s{2}}{\al}\p_{4\pi i\al}\int_{i\mathbb{R}}d\al'\,e^{4\pi i\al\al'}\chi_{\D'}(1|q)=\f1\al\p_{4\pi i\al}\chi_\D(1,\widetilde{q}\,)=\f1\tau\chi_\D(1,\widetilde{q}\,)
\end{multline}
This is in agreement with \eqref{modular transform toric} including the factor $\tau^{\D_e}$. Note that it is exactly this factor  that distinguishes the transformation laws \eqref{modular transform toric momentums} of $\D_e=0$ and $\D_e=1$ blocks leading to different kernels for the same blocks. We emphasize that in both cases the modular kernels have no poles at finite $\al$. This is expected in general: the analytic properties of the modular kernel must agree with those of conformal block.

\subsubsection{Single pole}
Let us now turn to single-pole conformal block \eqref{single pole explicit}. Recall that it arises when $b=i/\s{2}$ ($c=-2$) and $\D_e=3$ ($\D_e=\D_{1,3}$). We choose $\mu$ as $\mu=Q/2+\al_{1,3}=-3i/\s{2}$. Equation \eqref{mk equation} then specializes to
\begin{eqnarray}
-\f{\cos{2\pi \bar{b}\al}}{\sin{2\pi \bar{b} \al}}\left(e^{\f{\bar{b}}{2}\p_\al}-e^{-\f{\bar{b}}{2}\p_\al}\right)\mathcal{M_{\al\al'}}(\bar{\mu},\bar{b})=2\cos{2\pi \bar{b}\al'}\mathcal{M_{\al\al'}}(\bar{\mu},\bar{b})
\end{eqnarray}
where the notation $\bar{\mu}=-3i/\s{2}, \bar{b}=i/\s{2}$ is introduced. A possible solution is
\begin{eqnarray}
\mathcal{M_{\al\al'}}\left(\bar{\mu},\bar{b}\right)=2\s{2}\,\sin{4\pi\al\al'}\f{\sin{\s{2}\pi i\al'}}{\sin{\s{2}\pi i\al}}
\end{eqnarray}
while the normalization factor in this case becomes
\begin{eqnarray}
N_{\al\al'}\left(\bar{\mu},\bar{b}\right)=\f{\al'(1+8\al'^2)}{\al(1+8\al^2)}\f{\sin{\s{2}\pi i\al'}}{\sin{\s{2}\pi i\al}}
\end{eqnarray}
Again, disregarding the periodic trigonometric factor leads to the correct form of the full modular kernel
\begin{eqnarray}
\boxed{M_{\al\al'}\left(\bar{\mu},\bar{b}\right)=2\s{2}\,\f{\al'(1+8\al'^2)}{\al(1+8\al^2)}\sin{4\pi\al\al'}} \label{mk 1 pole sine toric}
\end{eqnarray}
Let us outline the verifying computation
\begin{multline}
\int_{i\mathbb{R}_+}d\al'\, M_{\al\al'}(\bar{\mu}, \bar{b})B_{\D'}(\D_e|q)\Big|_{\D_e=3,\, c=-2}=\\\f{\s{2}}{\al(1+8\al^2)}\int_{i\mathbb{R}}d\al'\,e^{4\pi i\al\al'}\al'(1+8\al'^2)\left(1+\f{1-E_2(q)}{-1-8\al'^2}\right)\chi_{\D'}(q)=\\\f{\s{2}}{\al(1+8\al^2)}\p_{4\pi i\al}(8\p_{4\pi i\al}^2+E_2(q))\int_{i\mathbb{R}}d\al'\,e^{4\pi i\al\al'}\chi_{\D'}(q)=\\\f{1}{\al(1+8\al^2)}\p_{4\pi i\al}(8\p_{4\pi i\al}^2+E_2(q))\chi_{\D'}(\widetilde{q}\,)=\\\f1{\tau^3}\chi_{\D'}(\widetilde{q}\,)\left(1-\f{1-\tau^2E_2(q)-6\tau/i\pi}{1+8\al^2}\right)=\\\f1{\tau^3}\chi_{\D'}(\widetilde{q}\,)\left(1+\f{1-(\tau^2E_2(q)+6\tau/i\pi)}{8\D}\right) \label{1 pole computation}
\end{multline}
which agrees with \eqref{modular transform toric momentums} including factor $\tau^\D_e$ and the anomalous transformation law of $E_2(q)$, see appendix \ref{special functions}. We emphasize again that the modular kernel as a function of $\al$ has the same poles as the conformal block.
\subsubsection{Two poles} \label{mk two poles toric sec}
As the last toric example we consider a two-pole conformal block \eqref{two pole explicit}. To realize this case we choose $b=i$ ($c=1$) and $\mu=2i$ ($\D_e=4$). We will start with the answer
\begin{eqnarray}
\boxed{M_{\al\al'}(2i,i)=2\s{2}\f{\al'^2(1+4\al'^2)}{\al^2(1+4\al^2)}\cos{4\pi \al\al'}} \label{mk 2 pole toric}
\end{eqnarray}
This result can be derived (up to a periodic factor) similarly to the collected examples. However, it is both simpler and more instructive to use the pattern observed previously and infer the answer. The principal part of the kernel is the Fourier-like contribution $\cos{4\pi\al\al'}$ (we will explain our preference of cosine over the sine shortly). The original conformal block \eqref{two pole explicit} has poles at $\D=0$ and $\D=1/4$ which at $b=i$ implies the second order pole at $\al=0$ and the simple poles at $\al=\pm1/2$. The minimal factor that accounts for these poles is $(\al^2(1+4\al^2))^{-1}$. We also have to include the factor $\al'^2(1+4\al'^2)$ (and numeric constant $2\s{2}$) to preserve the unit squaring property \eqref{mk unit square}. This gives expression \eqref{mk 2 pole toric}.

Another possible candidate satisfying the listed requirements is given by
\begin{eqnarray}
\widetilde{M}_{\al\al'}(2i,i)=2\s{2}\f{\al'^3(1+4\al'^2)}{\al^3(1+4\al^2)}\sin{4\pi \al\al'} \label{mk 2 pole toric false}
\end{eqnarray}
This expression behaves properly at $\al=0$ due to the zero of the sine function. However, it is clear beforehand that this kernel does not provide the correct modular transformation. Revisiting e.g. computation \eqref{1 pole computation} we see that the degree of a polynomial in $\al'$ is responsible for how many factors $\tau^{-1}$ appear in the resulting expression. Kernel \eqref{mk 2 pole toric false} would give $\tau^{-5}$ which is incorrect since there must be $\tau^{-\D_e}=\tau^{-4}$. In contrast, kernel \eqref{mk 2 pole toric} passes this last check. Similarly to the previous examples one can explicitly verify that the kernel is indeed correct. We will omit the computation.

Interestingly, this last criterion that we have formulated links the number of poles in a conformal block to the value of the external dimension. Namely, an $k$-pole block can either have $\D_e=2k$ or $\D_e=2k+1$. Or, from another angle, a finite-pole conformal block with the external dimension $\D_e$ (which is necessarily an integer, as shown earlier) has 
\begin{eqnarray}
k=\left \lfloor{\f{\D_e}{2}}\right \rfloor
\end{eqnarray}
poles. This conjectural relation holds for all the finite-pole blocks described in \cite{Beccaria:2016nnb}.

It is now clear how one guesses the modular kernel including the periodic factor undetermined by equation \eqref{mk equation}. In all the cases that we have checked the modular kernel is the Fourier kernel renormalized by the polynomial functions of $\al,\al'$. Everything beyond should be omitted. Moreover, these polynomials are fully defined by the poles of the conformal block and the value of $\D_e$. This specific structure of the modular kernels for the finite-pole blocks forms yet another consistent pattern which is tempting to promote to the general conjecture. And even more so since the structure is shared by the spheric blocks to which we now turn.
\subsection{Spheric}
We now briefly discuss the modular transformations of the spheric finite-pole blocks. In terms of the cross-ratio $x$ the relevant transformations are $x\to1-x$ and $x\to\f{x}{x-1}$. In terms of the spheric nome $q$ defined in \eqref{spheric nome} they take the usual form of the modular $S, T$ transformations. As in the toric case, the $T$ transformation acts simply by a phase factor and we only consider the $S$ transformation which we continue to call the modular transformation.

In contrast to the toric one-point function, the spheric correlator is invariant under the modular transformation so that conformal blocks satisfy \footnote{In fact, the modular transformation must be supplemented by swap of the external dimensions $\Delta_1\leftrightarrow\Delta_3$. We however omit this detail to lighten the notation. Also, we only consider the $\D_1=\D_3$ conformal blocks in the sequel.}
\begin{eqnarray}
B_{\D}(\D_i,c|\widetilde{q}\,)= \int\limits_{i\mathbb{R}_+}{d\al'} M_{\al\al'}(\D_i,c)B_{\D'}(\D_i,c|q) \label{modular transform spheric}
\end{eqnarray}
\subsubsection{No poles}
Up to permutations of the external dimensions there are two families of the no-pole conformal blocks \eqref{1 no pole family}, \eqref{2 no pole family}. Within both families the elliptic block equals one, but the prefactors \eqref{spheric prefactor} are different . The most important discrepancy is the power of the theta-function featuring in the prefactor
\begin{eqnarray}
D=\f{c-1}{2}-4(\D_1+\D_2+\D_3+\D_4) \label{theta power}
\end{eqnarray} 
For the first family $D=-1$ while for the second $D=-3$. Due to this distinction conformal blocks of these two types are transformed by the different modular kernels. One can check that the modular kernels for \eqref{1 no pole family}, \eqref{2 no pole family} are 
\begin{eqnarray}
\boxed{M_{\al\al'} = 2\s{2}\f{16^{-\al'^2}}{16^{-\al^2}}\cos{2\pi \al\al'}} \label{1 no pole mk}
\end{eqnarray}
and 
\begin{eqnarray}
\boxed{M_{\al\al'} = 2\s{2}\f{16^{-\al'^2}}{16^{-\al^2}}\f{\al'}{\al}\sin{2\pi \al\al'}} \label{2 no pole mk}
\end{eqnarray}
respectively. These are very reminiscent of the two modular kernels for the no-pole toric blocks \eqref{Fourier kernel}, \eqref{no pole sine kernel}. 
\subsubsection{Single pole}
As our final example we consider a one-pole spheric block \eqref{1 pole spheric} and apply the intuition inherited from the toric case to guess the modular kernel. Recall that the conformal block under discussion appears at $c=1$, $\D_1=\D_3=1/16$, $\D_2=\D_4=9/16$. As the answer we expect the kernel of the type \eqref{1 no pole mk} or \eqref{2 no pole mk} renormalized to account for the pole at $\D=0$ which at $c=1$ translates to the double pole at $\al=0$. There are two possible choices
\begin{eqnarray}
\boxed{M_{\al\al'}(\D_i,c) = 2\s{2}\f{16^{-\al'^2}}{16^{-\al^2}}\f{\al'^2}{\al^2}\cos{2\pi \al\al'}\label{1 pole mk spheric}}
\end{eqnarray}
or
\begin{eqnarray}
\widetilde{M}_{\al\al'}(\D_i,c) = 2\s{2}\f{16^{-\al'^2}}{16^{-\al^2}}\f{\al'^3}{\al^3}\sin{2\pi \al\al'} \label{1 pole mk false}
\end{eqnarray}
Similarly to the toric case, one can argue in advance that the latter option is not correct. Polynomial in $\al'$ of degree 3 will lead to the overall factor $\mathcal{T}^{-3}$ after integral \eqref{modular transform spheric} is evaluated. However, the power of the theta function from the prefactor \eqref{spheric prefactor} is $D=-5$. Due to the transformation law of $\theta_3$ (see appendix \ref{special functions}) this ensures the appearance of the multiplier $\mathcal{T}^{-2}$. This will match the result produced by kernel \eqref{1 pole mk spheric} but not \eqref{1 pole mk false}. Furthermore, in the spirit of e.g. equation \eqref{1 pole computation} one can check explicitly that kernel \eqref{1 pole mk spheric} provides the proper modular transformation.

In analogy with the toric case this observation suggests a relation between $D$ \eqref{theta power} and the number of poles $k$. Namely,

\begin{eqnarray}
k=\left\lfloor-\f{D+1}{4}\right\rfloor=\left\lfloor\D_1+\D_2+\D_3+\D_4-\f{c+1}{8}\right\rfloor
\end{eqnarray}

\section{Summary}
We have looked at the family of the exact conformal blocks recently found in \cite{Beccaria:2016nnb} from the standpoint of Zamolodchikov's formula. These special blocks are distinguished by the fact that they contain only a finite amount of poles in the $\D$-plane. Zamolodchikov's formula readily provides the necessary conditions for this phenomenon to happen: the central charge of the theory must be such that there is a pair of coincident degenerate dimensions \eqref{Kac zeroes} of the form $\D_{1,\pm(2M+1)}=\D_{2N+1,\pm1}$ for some $N, M\ge0$, see \eqref{general cond}. However, the analysis of these possibilities appears to be significantly more involved than is expected at the first glance.  

The simplest in this exotic family are the conformal blocks which contain no poles at finite $\D$. In classification of equation \eqref{general cond} they appear if either $N=0$ or $M=0$.
\begin{obs} \label{no pole obs}
At any value of the central charge toric conformal block with $\D_e=0$ or $\D_e=1$ contains no poles and is equal to the Virasoro character \eqref{character}. At any value of the central charge spheric conformal block with
\begin{eqnarray}
\D_1=\D_2=\D_3=\f{8+3b^{-2}+4 b^2}{16},\quad\D_4=\f{4+3b^{-2}}{16}
\end{eqnarray}
or
\begin{eqnarray}
\D_1=\D_2=\D_3=\f{8+3b^{-2}+4 b^2}{16},\quad\D_4=\f{3(4+b^{-2})}{16}
\end{eqnarray}
(or any combination of dimensions which is a permutation of the above) is free of poles and equal to function \eqref{spheric prefactor}. 
\end{obs}
For a particular pair $N, M\ge1$ conditions \eqref{general cond} restrict the central charge to a finite set of rational values of the form \eqref{minimal models c} with allowed values of $n, m\in\mathbb{Z}$ depending on $N, M$. We have gathered evidence that if $n,m$ are of the same sign ($c\le1$), which precisely corresponds to the central charges of the minimal models, we do obtain a finite-pole conformal blocks upon setting the external dimension equal to the coincident pair $\Delta_e=\D_{1,\pm(2M+1)}=\D_{2N+1,\pm1}$. On the other hand, when $n, m$ are of different signs ($c>1$) the corresponding conformal blocks can not be truncated to a finite pole number by any choice of the external dimension $\D_e$. Thus we make our 
\begin{obs} \label{central charge obs}
Finite-pole blocks only exist in theories with the central charges given by these of the minimal models
\begin{eqnarray}
c=1-6\f{(n-m)^2}{nm} \label{min mod c}
\end{eqnarray}
where $n,m$ are assumed to be coprime. This applies to both toric and spheric blocks.
\end{obs}
This conjecture is proven for toric blocks when $M=1$ and $N$ arbitrary (which is equivalent to $N=1$ and $M$ arbitrary) in appendix \ref{proof}. The proof is quite technical and bulky and we do not attempt to generalize it to $M>1$. Nevertheless, with computer assistance we have tested the hypothesis beyond $M=1$ up to and including $N\times M=6$. These tests are, however, not fully rigorous and proceed as follows. The conformal block $q$-expansion is computed to as many orders as possible in a given situation. For cases with $c>1$ one observes approximately linear growth of the number of poles with the order of the $q$-expansion: new poles appear almost every order. In contrast, for $c\le1$ the amount of poles settles at a constant value. The computational demands increase rapidly with the order and we typically operate within about $O(q^{10})$. One might argue that this accuracy is not enough to be convincing. However, in our view the overall coherence of results makes the conjecture very plausible.

\begin{obs} \label{toric dimensions obs}
For a given value of the central charge from equation \eqref{min mod c} there are infinitely many finite-pole blocks. Finite-pole truncation happens if the external dimension belongs to the following set
\begin{eqnarray}
\D_e\in\bigcup\limits_{\substack{n,m\in\mathbb{Z}_+\\p\in\mathbb{Z}\setminus \{0\}}}\Big\{(pn-1)(pm-1),\, p^2nm+pn-pm\Big\}
\end{eqnarray}
Moreover, we expect that this set covers all the finite-pole toric blocks. 
\end{obs}
Due to the larger number of parameters obtaining a similar classification within our approach in the spheric case would be quite involved and we do not attempt it here. The additional complexity is already evident at the example of the no-pole blocks described in observation \ref{no pole obs}.

\begin{obs} \label{q dependence obs}
Another remarkable property of the finite-pole blocks is that their $q$-dependence is expressible in terms of a finite number of the modular forms, while the maximum  weight of the appearing modular forms seems to be linearly related to the number of poles. Assuming such finite-weight modular ansatz one can compute these conformal blocks to all orders in $q$, see for example \eqref{single pole explicit}, \eqref{two pole explicit}, \eqref{1 pole spheric}.
\end{obs}
As explained in subsection the ansatz in necessary since \ref{single pole sec toric} Zamolodchikov's formula does not apparently allow to find this $q$-dependence to all orders or to explain why the maximum modular weight is finite. So in this regard our analysis adds little new to the results presented in \cite{Beccaria:2016nnb}.

Next, assuming that the conformal blocks under discussion belong to a CFT with continuous spectrum given by $\D\in\f{c-1}{24}+\mathbb{R}_+$ we have studied examples of the modular transformations explicitly constructing the corresponding kernels, see \eqref{modular transform toric momentums}, \eqref{modular transform spheric} for definitions. The results appear to form a clear pattern.
\begin{obs} \label{modular transform obs}
For the finite-pole block with poles at $\Delta=d_1,\dots, d_k$ the modular kernel is given by
\begin{eqnarray}
M_{\al\al'}=\frac{(\D(\al')-d_1)\dots(\D(\al')-d_k)}{(\D(\al)-d_1)\dots(\D(\al)-d_k)}m_{\al\al'},\qquad \D(\al)=\f{c-1}{24}-\al^2 \label{mk finite pole}
\end{eqnarray}
where
\begin{eqnarray}
m_{\al\al'}=\begin{cases}
2\s{2}\cos{4\pi \al\al'}, \quad\text{if}\quad \D_e-2k=0\\
2\s{2}\f{\al'}{\al}\sin{4\pi \al\al'}, \quad\text{if}\quad \D_e-2k=1
\end{cases} \label{mk options}
\end{eqnarray}
\end{obs}
Modulo minor differences the same seems to apply to the finite-pole spheric blocks. Namely, in the formula above it suffices to replace\footnote{This is basically a notational difference.} $\cos{4\pi \al\al'}\to\cos{2\pi \al\al'}$ and likewise for the sine function, introduce the factor $\f{16^{-\al'^2}}{16^{-\al^2}}$, and also use $D=\Delta_1+\dots+\D_4-\f{c-1}{2}$ in place of $\D_e$ to choose between two options in \eqref{mk options}. 

Formula \eqref{mk finite pole} has a clear interpretation. First of all, as shown in numerous works \cite{GMMpert, Nemkov1, Lerda, Billo:2013fi, GMMnonpert, Nemkov2, Nemkov:2015zha} the Fourier kernel-type factor is expected to be found in any modular kernel. Next, from the definitions \eqref{modular transform toric momentums}, \eqref{modular transform spheric} we expect the modular kernel to have the same analytic structure as conformal block as a function of the intermediate dimension $\D$. Polynomial factors in \eqref{mk finite pole} provide the simplest way to introduce these poles\footnote{We need both the numerator and the denominator to have property \eqref{mk unit square}.}. The minimal choice satisfying these criteria is then to pick the upper line of \eqref{mk options}. This possibility is often realized, see \eqref{Fourier kernel}, \eqref{mk 2 pole toric}, \eqref{1 pole mk spheric}. Note however, that the less obvious option in the second line of equation \eqref{mk options} also leads to the desired analytic properties. And it is realized in some cases, too \eqref{no pole sine kernel}, \eqref{mk 1 pole sine toric}, \eqref{2 no pole mk}. As stated in \eqref{mk options}, it possible to figure out the proper choice based on the number of poles $k$ and the value of $\D_e$. In formula \eqref{mk options} it is implicitly implied that $\D_e-2k$ can only be equal to either $0$ or $1$. This is quite non-trivial and deserves a separate
\begin{obs} \label{number of poles obs}
The number of poles $k$ in a finite-pole toric conformal block is directly related to the value of the external dimension
\begin{eqnarray}
k=\left\lfloor \f{\D_e}{2} \right\rfloor
\end{eqnarray}
where $\lfloor\dots\rfloor$ denotes the integer part. The counterpart of this relation for the spheric finite-pole blocks is
\begin{eqnarray}
k=\left\lfloor \D_1+\D_2+\D_3+\D_4-\f{c+1}{8} \right\rfloor
\end{eqnarray}
\end{obs}
As illustrated in subsection \ref{mk two poles toric sec} under the assumption that the proper kernel for the finite-pole block is the Fourier kernel renormalized by a polynomial both conjectures \ref{modular transform obs}, \ref{number of poles obs} can be proven. Moreover, there is an alternative way to determine the modular kernel based on solving certain equations. Unfortunately, for the finite-pole blocks these equations can not be solved uniquely, but only up to a certain non-polynomial multipliers. Assuming that such multipliers are absent one can independently derive formula \eqref{mk finite pole}. Eventually one has to consider the renormalization factor \eqref{toric vertex}. As a function of $\al$ it always has the same analytic structure as conformal block possibly up to an extra factor $\al'/\al$ which precisely distinguishes the two options in \eqref{mk options}. Then, specializing this factor to the particular $\D_e$ and $c$ and disregarding the non-polynomial contributions one arrives to \eqref{mk finite pole}.

\section{Discussion}
The core of the present paper is quite heavy on the technicalities. Each example separately seems to contain a lot of accidental features. However, when reconciled together cases analyzed form a surprisingly coherent and simple picture which we summarized as a collection of observations. Unfortunately, each of these observations is merely a conjecture. Yet, their overall consistency makes the whole structure much more solid. We expect that there are even more interrelations between these observations that we have revealed.

Most of these conjectures seem open for attacks via the toolbox of the present paper. Unfortunately, even in particular cases full proofs are lengthy and cumbersome. Nevertheless it is possible that with additional ingenuity and effort the conjectures could be tested within the current approach. On the other hand, it is tempting to look for an alternative point of view which would provide more suitable language for the problem. For example, the presence of the modular transformations for the finite-pole blocks suggests that they may be a part of some consistent CFT.

Finally, particular instances of the closed-form conformal blocks may be of an independent interest, regardless of whether or not they are embedded in a general theory. We expect the techniques of the present paper to be applicable in other situations where the recursion of Zamolodchikov's type is available, for example for the superconformal blocks \cite{Suchanek:2010kq, Hadasz:2012im}. Also, it would be interesting to relate the finite-pole blocks to the general context of the non-perturbative conformal blocks initiated in \cite{Itoyama:2014dya}.

\subsection*{Acknowledgements}
The author is grateful to Alexandra Anokhina and Gleb Aminov for useful discussions, to Semeon Arthamonov for help with the manuscript and to Alexei Morozov and Andrey Mironov for their guidance. The work is partly supported by grants RFBR 16-01-00291, RFBR 16-32-00920-mol-a, RFBR 15-51-52031-NSC-a, RFBR 16-51-53034-GFEN, RFBR 15-51-50034-YaF.

\appendix
\section{Special functions} \label{special functions}
The Dedekind eta-function naturally appears as a part of toric conformal block
\begin{eqnarray}
\eta(q)=q^{\f1{24}}\prod_{n=1}^\infty(1-q^n)
\end{eqnarray}
In the computations of the main text the second and the fourth Eisenstein series appear, which we define as
\begin{eqnarray}
E_2(q)=1-24\sum_{n=1}\f{nq^n}{1-q^n},\qquad
E_4(q)=1+240\sum_{n=1}\f{n^3q^n}{1-q^n}
\end{eqnarray} 
Besides, the elliptic theta functions are used
\begin{eqnarray}
\theta_2(q)=\sum_{n\in\mathbb{Z}}q^{(n-1/2)^2},\quad \theta_3(q)=\sum_{n\in\mathbb{Z}}q^{n^2},\quad \theta_4(q)=\sum_{n\in\mathbb{Z}}(-1)^nq^{n^2}
\end{eqnarray}
Here the nome $q$ is assumed to be $q=e^{2\pi i \tau}$. It is the modular properties of these functions under transformation $\tau\to-1/\tau$ that matter for our purposes. These properties are summarized as follows
\begin{align}
& \eta(\widetilde{q}\,)=\s{-i\tau}\eta(q)\\
& E_2(\widetilde{q}\,)=\tau^2E_2(q)+6\tau/i\pi,\qquad E_4(\widetilde{q}\,)=\tau^4E_4(q)\\
& \theta_2(\widetilde{q}\,)=\s{-i\tau}\theta_4(q), \quad
 \theta_4(\widetilde{q}\,)=\s{-i\tau}\theta_2(q), \quad
 \theta_3(\widetilde{q}\,)=\s{-i\tau}\theta_3(q) 
\end{align}
where $\tilde{q}=e^{-2\pi i/\tau}$.

In the convenient renormalization of conformal blocks \eqref{toric vertex} the so-called double Gamma function $\Gamma_b(z)$ appears. It can be defined as the analytic continuation of the integral representation
\begin{eqnarray}
\Gamma_b(z)=\int_0^\infty \f{dt}{t}\left(\f{e^{-zt}-e^{-Qt/2}}{(1-e^{-bt})(1-e^{-b^{-1}t})}-\f{(Q-2z)^2}{8e^t}-\f{Q-2z}{2t}\right),\qquad Q=b+b^{-1}
\end{eqnarray}
Note the self-duality property $\Gamma_b(z)=\Gamma_{b^{-1}}(z)$. $\Gamma_b(z)$ is a meromorphic function of $z$ with no zeros and poles located at $z=-rb-sb^{-1}$ for $r,s\ge0$ so that
\begin{eqnarray}
\Gamma_b(z)\propto \prod_{r,s\ge0}\f1{z+rb+sb^{-1}}
\end{eqnarray}
Within the scope of the present paper its defining property is the difference equation
\begin{eqnarray}
\Gamma_b(z+b)=\Gamma_b(z)\f{\s{2\pi}b^{bz-1/2}}{\Gamma(bz)}
\end{eqnarray}
\section{On central charges of finite-pole blocks} \label{proof}
We have conjectured that among solutions of \eqref{general cond} only those with $c\le1$ lead to the finite-pole blocks. We now prove this for $M=1$ and all $N\ge1$\footnote{This is of course equivalent to $N=1, M\ge1$ by the symmetry $b\leftrightarrow b^{-1}$ which swaps the indices of the degenerate dimensions $\D_{r,s}\to \D_{s,r}$.}. The proof closely parallels the analysis of  section \ref{single pole sec toric}. The subtleties arise since certain terms in Zamolodchikov's formula might be singular when the central charge takes the values of interest. Hence, the  corresponding limits must be analyzed with care.

For $M=1$ the external dimension can be either $\D_{1,3}$ or $\D_{1,-3}$. For brevity we denote both possibilities by $\bar{\D}_e=\D_{1,\pm3}$. Upon setting $\D_e=\bar{\D}_e$ all the coefficients $R_{r\ge1,s\ge2}(\D_e,c)$ vanish and the general recurrence relation \eqref{Zamolodchikov's elliptic} is reduced to
\begin{eqnarray}
H_{\D}(\bar{\D}_{e},c|q)=1+\sum_{r\ge1}\f{R_{r,1}(\bar{\D}_{e},c)}{\D-\D_{r,1}}q^rH_{\D_{r,-1}}(\bar{\D}_e,c|q) \label{reduced recursion}
\end{eqnarray}
\subsection*{$c>1$}
Assume that a given $c>1$ corresponds to $\al_{n,m}=0$ ($nm<0$). We shall now demonstrate that in the limit $\al_{n,m}\to0$ an infinite amount of terms survives in \eqref{reduced recursion}.  Indeed, let $r$ be such that the following limit is non-vanishing
\begin{eqnarray}
\lim\limits_{\al_{n,m}\to0}R_{r,1}(\bar{\D}_e,c)H_{\D_{r,-1}}(\bar{\D}_e,c|q)\neq0
\end{eqnarray} 
This means that there is at least one pole at $\D=\D_{r,1}$ in the conformal block under discussion. Then, one can show that there exist $r'>r$  such that $R_{r',1}(\bar{\D}_e,c)H_{\D_{r',-1}}(\bar{\D}_e,c|q)$ is also non-vanishing in this limit and therefore contributes a pole at $\D=\D_{r',1}$. Hence, by induction, the amount of poles is infinite.

To demonstrate this let us first substitute $\D=\D_{r',-1}$ in \eqref{reduced recursion}
\begin{eqnarray}
H_{\D_{r',-1}}(\bar{\D}_e,c|q)=1+\sum_{r\ge1}\f{R_{r,1}(\bar{\D}_e,c)}{\D_{r',-1}-\D_{r,1}}q^rH_{\D_{r,-1}}(\bar{\D}_e,c|q) \label{reduced reccurence part}
\end{eqnarray}
The conformal block in the l.h.s might be singular due to singularity of the denominator in the r.h.s. This happens when $\D_{r',-1}=\D_{r,1}$ which at $c>1$ implies
\begin{eqnarray}
\al_{r'-r,-2}=0
\end{eqnarray} 
In the case at hand ($M=1$) there four possible choices leading to $c>1$, see \eqref{general cond}
\begin{eqnarray}
\al_{N,-1}=0,\qquad \al_{N,-2}=0,\qquad \al_{N+1,-1}=0,\qquad \al_{N+1,-2}=0
\end{eqnarray}
Choosing 
\begin{eqnarray}
r'=r+2N,\qquad r'=r+N,\qquad r'=r+2N+2,\qquad r'=r+N+1
\end{eqnarray}
in each of these cases respectively we satisfy condition $\al_{r'-r,-2}=0$. Hence the corresponding conformal blocks $H_{\D_{r',-1}}(\bar{\D}_e,c)$ as functions of the central charge have simple poles.

It remains to demonstrate the accompanying factors $R_{r',1}(\bar{\D}_e,c)$ do not have multiple zeros at these points. Let us consider
\begin{eqnarray}
R_{r,1}(\D_{1,3},c)=\f{\al_{r,1}}{Q}\prod_{n=0}^{r-1}\f{(\D_{1,3}-\D_{2n+1,1})(\D_{1,3}-\D_{2n+1,-1})}{\D'_{2n+1,1}\D_{2n+1,-1}} \label{reduced R}
\end{eqnarray}
This expression in fact never has multiple zeros. Indeed, the numerator vanishes if either $\D_{1,3}=\D_{2n+1,1}$ or $\D_{1,3}=\D_{2n+1,-1}$ which at $c>1$ implies $\al_{n,-1}=0$ or $\al_{n,-2}=0$. However, at $\al_{n,-1}$ the denominator is also singular due to vanishing of $\D_{2n+1,-1}$. Hence, one of the zeros that may occur in the numerator always comes together with the zero in the denominator. We conclude that the expression \eqref{reduced R} never features multiple zeros. Similar arguments work for $\D_e=\D_{1,-3}$ and we omit them. This finishes our treatment of the $c>1$ case.

\subsection*{$c\le1$}
Let us re-examine equation \eqref{reduced reccurence part} at $c\le1$. Coefficient conformal blocks $H_{\D_{r',-1}}(\D_e,c)$ are singular when there are collision of dimensions of the type $\D_{r',-1}=\D_{r,1}$ which implies $\al_{r'+r,0}=0$ or $\al_{r'-r,-2}=0$. Since $r'>r>0$ neither of these equations has solutions at $c\le1$ and therefore no singularities akin to $c>1$ case appear.

A remaining caveat is that coefficients $R_{r,1}(\D_e,c)$ could develop singularities of their own. In full analogy with the $c>1$ case one can show that this is not the case, since the zeros of the denominator are shared by the numerator. Therefore, the amount of poles in the $c\le1$ case is indeed finite.

\bibliographystyle{utcaps_edited}
\bibliography{bibfile,revtex-custom}

\end{document}